# Tuning of the temperature dependence of the resonance frequency shift in atomically thin mechanical resonators with van der Waals heterojunctions

Taichi Inoue, Yuta Mochizuki, Kuniharu Takei, Takayuki Arie, and Seiji Akita[1]

Department of Physics and Electronics, Osaka Prefecture University, Sakai 599-8531, Japan

Atomically thin two-dimensional (2D) mechanical resonators should realize highly sensitive force sensors and high performance nano-electro-mechanical systems due to their excellent electrical and mechanical properties. However, practical applications require stability of the resonance frequencies against temperature. Here, we demonstrate the manipulation of the thermal expansion coefficients (TECs) by creating a van der Waals heterojunction using graphene and $MoS_2$, which have opposite signs of TECs. Our method greatly suppresses the apparent TEC of the 2D heterojunction to 1/3 of the monolayer graphene without the detraction of the tunability of the resonance frequency by electrostatic attraction.

---

[1] e-mail: akita@pe.osakafu-u.ac.jp

Atomically thin two-dimensional (2D) materials such as graphene and transition metal dichalcogenides (TMDC) are attractive for nano-electro-mechanical systems (NEMS) due to their excellent electrical and mechanical properties.[1-5] In particular, mechanical resonators (MRs) composed of atomically thin 2D materials[5-19] have widely been investigated. Owing to their extremely light weight, they show a very high resonance frequency, which is beneficial for force sensor applications. In addition, oscillators[16, 20] with a tunable resonance frequency are candidates for the base-clock element in digital circuits. Toward computing applications, atomically thin 2D MRs exhibit energy transfer across the oscillation modes[21, 22], but depending on the temperature, the thermal stress on atomically thin 2D membranes of MR modulates the resonance frequency.

This modulation results in unexpected fluctuations in the resonance frequency. These temperature induced fluctuations prevent stable operations of MR devices. For instance, strong perturbations make accurately measuring the force using atomically thin 2D MR difficult. Although suppression of the temperature dependence of the resonance frequency at a specific gate bias condition has been reported,[18] the tunability of the resonance frequency is limited in this case. Thus, the tuning of the resonance frequency should be independent of the bias condition.

Thermal stress is a general problem not only in atomically thin 2D MR but also in conventional bulk NEMS. Temperature fluctuations easily affect the stability of NEMS-MR. To suppress the thermal stress, the thermal expansion coefficient (TEC) of MRs, which consist of bulk materials, have been tuned by chemical doping or introducing defects in the materials.[23-25] However, both approaches may degrade the superb electrical and mechanical properties of atomically thin 2D materials. Especially, the introduction of defects degrades the quality factor of the resonance, which is the most important figure of merit for MR, even in nanoscale resonators[26]. Hence, establishing simple techniques to tune TEC for atomically thin materials is desirable.

Tuning of the electrical properties in atomically thin 2D materials is widely investigated by stacking different kinds of atomically thin 2D materials[27-31] to form a van der Waals (vdW) heterojunction. In a vdW, the stacked 2D materials have different electrical properties such as electronic band structures. In this regard, we expect that a vdW heterojunction can tune the mechanical properties of atomically thin 2D materials[17]. Herein we demonstrate that stacking graphene and $MoS_2$, which have different signs of TECs, suppresses the temperature dependence of the frequency shift in atomically thin 2D MR.

Figure 1a schematically illustrates the $MoS_2$/graphene stacked MR. First, metal electrodes consisting of Cr/Au (5 nm/30 nm) as a support of the graphene drum were fabricated on a heavily doped n-type Si substrate (< 0.02 $\Omega$cm) with a 300 nm-thick $SiO_2$ layer using conventional photolithography. After the monolayer graphene was transferred onto the substrate using polymethyl methacrylate, the sample was trimmed using oxygen plasma etching to form the drum, where the graphene was synthesized using low-pressure chemical vapor deposition at 1000 °C and Cu foil as a catalyst.[32, 33] $MoS_2$ flakes were transferred on graphene by the conventional polydimethylpolysiloxane (PDMS) gel stamp method,[31] where bulk $MoS_2$ was mechanically exfoliated with a thermal release tape and subsequently transferred onto the PDMS gel stamp.

To form the drum-type graphene/$MoS_2$ (G/$MoS_2$)-MR suspended by metal electrodes, the $SiO_2$ layer underneath the graphene drum was etched using buffered HF, where the metal electrodes were used as the metal mask for etching.[15] Trenches on both sides of the drum are necessary for uniform etching of the $SiO_2$ layer as they serve as an evacuation channel for the air between the G/$MoS_2$ drum and the substrate while measuring the resonance characteristics. The samples were finally dried using supercritical drying to prevent sticking of the suspended structure induced by the surface tension of water.

Figure 1b shows a scanning electron microscope (SEM) image of the prepared sample. The heavily doped Si substrate acts as a back gate. The gap between the G/MoS$_2$ membrane and the substrate is 300 nm, which corresponds to the SiO$_2$ layer thickness. The diameter of the drum part of the G/MoS$_2$-MR is 6 μm. As for control, we prepared a graphene-only MR (G-MR) as well as a G/MoS$_2$-MR. Figure 1c and 1d show the Raman spectra for MoS$_2$ and graphene after the fabrication process, respectively. There are three MoS$_2$ layers.[34] It should be noted that the D band peak for graphene observed at 1350 cm$^{-1}$ is hardly observed, indicating that the fabrication process induces few defects.

For the resonance property measurements, the AM modulation down mixing method[35] was used. As shown in Fig. 2a, the graphene mainly acts as the FET channel. All measurements were performed in a vacuum (less than 10$^{-3}$ Pa) after annealing at 150 °C for 2 hours. Figure 2b shows the DC transfer characteristic of the G/MoS$_2$-MR without AC modulation at a source-drain voltage, $V_{sd}^{DC}$ = 5 mV. The gate voltage, $V_{gs}$, successfully controls the drain current, $I_{ds}^{DC}$, even for the suspended graphene/MoS$_2$ stacking channel after the sample preparation process. It should be noted that the drain currents for MoS$_2$ FETs with a similar structure (less than 100 nA) show much smaller channels than that for graphene under $V_{sd}^{DC}$ = 5 mV. Thus, the current passing through the graphene layer mainly contributes to the measured $I_{ds}^{DC}$. To prevent the G/MoS$_2$ membrane from sticking onto the substrate, the gate bias is limited to less than 5 V.

Figure 2c shows the frequency response curves of the G-MR under various $V_{gs}$ when $V_{sd}^{AC}$ = 7 mV$_{rms}$ without a DC component is applied at a modulation frequency of 1 kHz and a modulation depth of 99%. The small arrows denote the resonance frequency, $f_0$. As $V_{gs}$ increases, $f_0$ shifts toward a higher frequency due to the increased internal strain of the graphene membrane.

Figure 2d shows the frequency response curves of G/MoS$_2$-MR. Similar to G-MR in Fig. 2c, the resonance frequency shifts toward a higher frequency with increasing $V_{gs}$. The mixed down current ($I_{MOD\_AM}$) measured for G/MoS$_2$-MR is attributed to the current passing through the graphene layer due to the higher conductivity of graphene than that of MoS$_2$ as mentioned above.

These results demonstrate that an electrical method can measure the resonance curves of G/MoS$_2$-MR. The frequency responses for both samples are well fitted to the linear response curve calculated with a parameter for the phase differences between $V_{ds}^{AC}$ and $I_{AM\_MOD}$ for all $V_{gs}$. Thus, we conclude that the frequency responses observed in these experiments are in the linear response regime. It should be noted that the low cutoff frequency (less than 1 MHz) of the MoS$_2$ FET in the experimental conditions prevents measurements of the frequency response of MoS$_2$-only MR.

Figures 3a and 3b show the $V_{gs}$ dependences of $f_0$ for the G-MR and G/MoS$_2$-MR measured under various temperatures, $\Delta T$, above room temperature (295 K) in a vacuum, respectively. For both samples, the resonance frequencies increase with increasing $V_{gs}$ at all temperatures. At $V_{gs} < 4$ V for G-MR, $f_0$ depends on $\Delta T$, and increases with increasing $\Delta T$. This is most likely due to the negative TEC of graphene.[18] At $V_{gs} = 4.5$ V, the $\Delta T$ dependence of $f_0$ almost vanishes. Further increasing $V_{gs}$ corresponds to the higher strain and results in the opposite temperature dependence of $f_0$. A similar behavior has been already reported in the low temperature regime as discussed later.[18] In contrast to G-MR, the temperature dependence of $f_0$ for G/MoS$_2$-MR is greatly suppressed, even under a lower strain ($V_{gs} < 4$ V). However, like G-MR, the temperature dependence of G/MoS$_2$-MR seems to decrease at higher $V_{gs}$.

To clarify the temperature dependence, Figs. 4a and 4b plot $\Delta f_0/f_0$ against $\Delta T$ for G-MR and G/MoS$_2$-MR, respectively, where $\Delta f_0$ is the resonance frequency shift measured at room temperature for each $V_{gs}$. The temperature dependence of $\Delta f_0/f_0$ for G-MR strongly depends on $V_{gs}$ at $V_{gs} < 4$ V, as mentioned in Fig. 3a. At $V_{gs} = 4$ V, the temperature dependence almost disappears and becomes negative at higher $V_{gs}$. Thus, the temperature dependence $\Delta f_0/f_0$ can be tuned by $V_{gs}$. However, it is hard to tune the resonance frequency by $V_{gs}$ with remaining the minimum temperature dependence, limiting the practical applications for a tunable oscillator. In the case of G/MoS$_2$-MR shown in Fig. 4b, the temperature dependence is greatly suppressed to less than 2% for $\Delta T$ = 10 K, even at low $V_{gs}$. This is one order of magnitude smaller than that for G-MR. Hence, the resonance frequency can be tuned by $V_{gs}$ with small fluctuations in the resonance frequency against the temperature.

Figure 4c summarizes the temperature coefficient of $\Delta f_0/f_0$ for the respective $V_{gs}$ estimated from Figs. 4a and 4c. As mentioned before, the temperature coefficient of $\Delta f_0/f_0$ decreases with increasing $V_{gs}$, which corresponds to the increase in the internal strain. The reported values of TEC for graphene $\alpha_{gra}$[18, 36] and MoS$_2$ $\alpha_{MoS2}$[4] at room temperature are in the range of $-3 \sim -8 \times 10^{-6}$ K$^{-1}$ and $\sim 5 \times 10^{-6}$ K$^{-1}$, respectively. Although the opposite signs of TECs cause the membrane to bend, this thermal stress is relaxed due to the presence of many wrinkles, as observed in Fig. 1b. In addition, the MoS$_2$(3L)/graphene(1L)-MR with a 6-μm diameter examined here can be treated as a membrane,[2, 19] where the membrane tension is the dominant parameter in the resonance frequency. If this model is valid, membrane bending has a limited impact on the temperature dependence. Here, we consider the apparent TEC $\alpha_{app}$ for the stacked membrane, which is given as

$$\alpha_{app} = \frac{E_{gra}\alpha_{gra}n_{gra} + E_{MoS_2}\alpha_{MoS_2}n_{MoS_2}}{E_{gra}n_{gra} + E_{MoS_2}n_{MoS_2}}. \tag{1}$$

where $E_{gra}$ and $E_{MoS2}$ are the monolayer Young's moduli, and $n_{gra}$ and $n_{MoS2}$ are the number of layers of graphene and MoS$_2$, respectively. In our experiments, the apparent TECs are in the range of $-1.8 \sim 1.3 \times 10^{-6}$ K$^{-1}$ owing to the compensation of $\alpha_{gra}$ and $\alpha_{MoS2}$ upon stacking, which results in a smaller temperature dependence of the resonance frequency of G/MoS$_2$-MR at a wider range of $V_{gs}$.

To better understand the observed differences between G-MR and G/MoS$_2$-MR, the observed temperature and $V_{gs}$ dependences of $f_0$ are investigated based on a model for the rectangular G-MR proposed by Singh et al.[18] To apply this rectangular model to the circular drum-shaped MR, we assumed an extremely simplified model where that the length and width of the rectangular shape are on the order of the diameter, $d_0$. In the case of the MoS$_2$(3L)/graphene(1L)-MR examined here, it can be treated as a membrane[12), 13)], where the tension of the membrane is the dominant parameter in the resonance frequency. In this model, the fundamental resonance frequency is given by

$$f_0(V_{gs}, \Delta T) = \frac{1}{2\pi}\left(\frac{\pi^2 \Gamma}{d_0^3 \rho t} - \frac{\epsilon_0 (V_{gs})^2}{\rho t d_t^3}\right)^{1/2}, \quad (2)$$

where $\Gamma$ is the tension of the membrane in plane, $\rho$ is the apparent mass density of the resonator, and $t$ is the thickness. $\epsilon_0$ is the electric constant and $d_t$ is the gap between the membrane and the substrate for the electrostatic attraction. The apparent mass density is given by $\rho = \frac{\rho_{gra}t_{gra} + \rho_{MoS_2}t_{MoS_2}}{t_{gra} + t_{MoS_2}}$, where $\rho_{gra}$ and $\rho_{MoS2}$ are the mass densities, and $t_{gra}$ and $t_{MoS2}$ thicknesses of graphene and MoS$_2$, respectively. The tension, $\Gamma(V_{gs}, \Delta T)$, is expressed as

$$\Gamma(V_{gs}, \Delta T) \approx \Gamma_0 + \frac{Et}{\Gamma_0^2}\frac{\epsilon_0^2 d_0^5}{96 d_t^4} V_{gs}^4, \quad (3)$$

where

$$\Gamma_0(\Delta T) \approx \Gamma_{00}\left(1 - \frac{Ed_0 t}{\Gamma_{00}} \alpha_{eff} \Delta T\right). \quad (4)$$

$\Gamma_{00}$ is the initial tension at $\Delta T = 0$ and $V_{gs} = 0$. $E$ is the apparent Young's modulus of the membrane, and $\alpha_{eff}$ is the effective TEC of entire device, including the membrane, substrate, and electrodes. We further assumed that graphene and MoS$_2$ have the same Poisson's ratio. Thus, the apparent Young's modulus of G/MoS$_2$-MR is $E \approx \frac{E_{gra}n_{gra} + E_{MoS_2}n_{MoS_2}}{n_{gra} + n_{MoS_2}}$. Finally, inserting eqs. 3 and 4 into eq. 2 gives

$$f_0(V_{gs}, \Delta T) = f_{00}\left(1 - \frac{b\alpha_{eff}}{\Gamma_{00}}\Delta T + \frac{a}{\Gamma_{00}^3\left(1 - \frac{b\alpha_{eff}}{\Gamma_{00}}\Delta T\right)^2}V_{gs}^4 - \frac{d_0^3 \epsilon_0}{\pi^2 \Gamma_{00} d_t^3}V_{gs}^2\right)^{1/2}, \quad (5)$$

where $f_{00} = \frac{1}{2}\left(\frac{\Gamma_{00}}{d_0^3 \rho t}\right)^{1/2}$, $a = \frac{Et\epsilon_0^2 d_0^5}{96 d_t^4}$, and $b = E d_0 t$. We fitted the experimental results to eq. 5 (Figs. 3 and 4, solid lines). Note that because we assumed the drum is rectangular, the geometrical factors considered in eq. 5 are invalid.

Even under this rough assumption, both the $V_{gs}$ and $\Delta T$ dependences qualitatively agree well with eq. 5, except for G/MoS$_2$-MR at $\Delta T = 0$ (Figs. 3 and 4). From the fitting parameters, we roughly evaluate the apparent TECs of G-MR and G/MoS$_2$-MR to be $-1.6\times10^{-6}$ K$^{-1}$ and $-5.9\times10^{-7}$ K$^{-1}$, respectively. Table I summarizes the parameters used in this fitting. As expected, the apparent TEC of G/MoS$_2$-MR is greatly suppressed to 1/3 of G-MR by creating a vdW heterojunction (Fig. 4). In this way, the temperature dependence of the resonance frequency of the atomically thin MR is successfully manipulated without tuning the resonance frequency by the gate bias.

This study investigated the apparent TEC of atomically thin membranes using an electrostatically actuated MR from their mechanical resonance frequencies. Stacking two different atomic layers, graphene and MoS$_2$, which have the opposite signs of TECs, suppresses the apparent TEC of the atomically thin drum-type MR. Due to the reduction in the apparent TEC, the temperature dependence of the resonance frequency shift decreases without the detraction of the tunability of the resonance frequency by $V_{gs}$. Consequently, the resonance frequency shift shrinks from 0.25 to 0.15% K$^{-1}$ as the electrostatic attraction increases. We believe that this strategy, which manipulates TEC, will realize additional applications of atomically thin MR and eventually achieve highly accurate NEMS sensors with an improved temperature stability.

**Acknowledgements**

This work was partially supported by KAKENHI Grant Numbers 16H00920, 16K14259, 16H06504, and 17H01040.

**Author Contributions**

T.I. and S.A. conceived and designed the experiments, led the research, and wrote the paper. Y.M. contributed to sample preparation. K.T. and T.A. contributed to data analysis. All authors discussed the results and assisted in manuscript preparation.

**Competing interests:** The authors declare that they have no competing interests.

**References**

1. Lee, C.; Wei, X.; Kysar, J. W.; Hone, J. *Science* **2008,** 321, (5887), 385-388.
2. Bolotin, K. I.; Sikes, K. J.; Jiang, Z.; Klima, M.; Fudenberg, G.; Hone, J.; Kim, P.; Stormer, H. L. *Solid State Commun* **2008,** 146, (9), 351-355.
3. Castellanos-Gomez, A.; Poot, M.; Steele, G. A.; van der Zant, H. S. J.; Agraït, N.; Rubio-Bollinger, G. *Adv. Mater.* **2012,** 24, (6), 772-775.
4. Gan, C. K.; Liu, Y. Y. F. *Phys. Rev. B* **2016,** 94, (13), 134303.
5. Lee, J.; Wang, Z.; He, K.; Shan, J.; Feng, P. X. L. *ACS Nano* **2013,** 7, (7), 6086-6091.
6. Bunch, J. S.; van der Zande, A. M.; Verbridge, S. S.; Frank, I. W.; Tanenbaum, D. M.; Parpia, J. M.; Craighead, H. G.; McEuen, P. L. *Science* **2007,** 315, (5811), 490-493.
7. van der Zande, A. M.; Barton, R. A.; Alden, J. S.; Ruiz-Vargas, C. S.; Whitney, W. S.; Pham, P. H.; Park, J.; Parpia, J. M.; Craighead, H. G.; McEuen, P. L. *Nano Lett.* **2010,** 10, (12), 4869-73.
8. Barton, R. A.; Storch, I. R.; Adiga, V. P.; Sakakibara, R.; Cipriany, B. R.; Ilic, B.; Wang, S. P.; Ong, P.; McEuen, P. L.; Parpia, J. M.; Craighead, H. G. *Nano Lett.* **2012,** 12, (9), 4681-6.
9. Croy, A.; Midtvedt, D.; Isacsson, A.; Kinaret, J. M. *Phys. Rev. B* **2012,** 86, (23), 235435.
10. Jiang, J. W.; Park, H. S.; Rabczuk, T. *Nanotechnology* **2012,** 23, (47), 475501.
11. Takamura, M.; Furukawa, K.; Okamoto, H.; Tanabe, S.; Yamaguchi, H.; Hibino, H. *Jpn. J. Appl. Phys.* **2013,** 52, (4S), 04CH01.
12. Miao, T.; Yeom, S.; Wang, P.; Standley, B.; Bockrath, M. *Nano Lett.* **2014,** 14, (6), 2982-7.
13. Takamura, M.; Okamoto, H.; Furukawa, K.; Yamaguchi, H.; Hibino, H. *J. Appl. Phys.* **2014,** 116, (6), 064304.
14. Chen, C.; Deshpande, V. V.; Koshino, M.; Lee, S.; Gondarenko, A.; MacDonald, A. H.; Kim, P.; Hone, J. *Nat. Phys.* **2016,** 12, (3), 240-244.
15. Inoue, T.; Anno, Y.; Imakita, Y.; Takei, K.; Arie, T.; Akita, S. *ACS Omega* **2017,** 2, (9), 5792 - 5797.
16. Lee, J.; Wang, Z.; He, K.; Yang, R.; Shan, J.; Feng, P. X.-L. *Science Advances* **2018,** 4, (3), eaao6653.17.    Ye, F.; Lee, J.; Feng, P. X. *Nanoscale* **2017,** 9, (46), 18208-18215.
18. Singh, V.; Sengupta, S.; S. Solanki, H.; Dhall, R.; Allain, A.; Dhara, S.; Pant, P.; M. Deshmukh, M. *Nanotechnology* **2010,** 21, (16), 165204.
19. Castellanos-Gomez, A.; van Leeuwen, R.; Buscema, M.; van der Zant, H. S.; Steele, G. A.; Venstra, W. J. *Adv. Mater.* **2013,** 25, (46), 6719-23.
20. Chen, C.; Lee, S.; Deshpande, V. V.; Lee, G.-H.; Lekas, M.; Shepard, K.; Hone, J. *Nat. Nanotechnol.* **2013,** 8, 923.
21. De Alba, R.; Massel, F.; Storch, I. R.; Abhilash, T. S.; Hui, A.; McEuen, P. L.; Craighead, H. G.; Parpia, J. M. *Nat. Nanotechnol.* **2016,** 11, 741.
22. Mathew, J. P.; Patel, R. N.; Borah, A.; Vijay, R.; Deshmukh, M. M. *Nat. Nanotechnol.* **2016,** 11, 747.

23. Chen, J.; Hu, L.; Deng, J.; Xing, X. *Chemical Society Reviews* **2015,** 44, (11), 3522-3567.
24. Romao, C. P.; Perras, F. A.; Werner-Zwanziger, U.; Lussier, J. A.; Miller, K. J.; Calahoo, C. M.; Zwanziger, J. W.; Bieringer, M.; Marinkovic, B. A.; Bryce, D. L.; White, M. A. *Chem Mater* **2015,** 27, (7), 2633-2646.
25. López-Polín, G.; Ortega, M.; Vilhena, J. G.; Alda, I.; Gomez-Herrero, J.; Serena, P. A.; Gomez-Navarro, C.; Pérez, R. *Carbon* **2017,** 116, 670-677.
26. Sawaya, S.; Akita, S.; Nakayama, Y. *Nanotechnology* **2007,** 18, (3), 035702.
27. Roy, K.; Padmanabhan, M.; Goswami, S.; Sai, T. P.; Ramalingam, G.; Raghavan, S.; Ghosh, A. *Nat. Nanotechnol.* **2013,** 8, 826.
28. Wang, F.; Wang, Z.; Xu, K.; Wang, Q.; Huang, Y.; Yin, L.; He, J. *Nano Lett.* **2015,** 15, (11), 7558-66.
29. Furchi, M. M.; Pospischil, A.; Libisch, F.; Burgdörfer, J.; Mueller, T. *Nano Lett.* **2014,** 14, (8), 4785-4791.
30. Deng, Y.; Luo, Z.; Conrad, N. J.; Liu, H.; Gong, Y.; Najmaei, S.; Ajayan, P. M.; Lou, J.; Xu, X.; Ye, P. D. *ACS Nano* **2014,** 8, (8), 8292-8299.
31. Fang, H.; Battaglia, C.; Carraro, C.; Nemsak, S.; Ozdol, B.; Kang, J. S.; Bechtel, H. A.; Desai, S. B.; Kronast, F.; Unal, A. A.; Conti, G.; Conlon, C.; Palsson, G. K.; Martin, M. C.; Minor, A. M.; Fadley, C. S.; Yablonovitch, E.; Maboudian, R.; Javey, A. *Proceedings of the National Academy of Sciences* **2014,** 111, (17), 6198-6202.
32. Anno, Y.; Takei, K.; Akita, S.; Arie, T. *Phys. Status. Solidi.-RRL* **2014,** 8, (8), 692-697.
33. Li, X.; Cai, W.; An, J.; Kim, S.; Nah, J.; Yang, D.; Piner, R.; Velamakanni, A.; Jung, I.; Tutuc, E.; Banerjee, S. K.; Colombo, L.; Ruoff, R. S. *Science* **2009,** 324, (5932), 1312-1314.
34. Li, H.; Zhang, Q.; Yap Chin Chong, R.; Tay Beng, K.; Edwin Teo Hang, T.; Olivier, A.; Baillargeat, D. *Adv. Funct. Mater.* **2012,** 22, (7), 1385-1390.
35. Sazonova, V.; Yaish, Y.; Ustunel, H.; Roundy, D.; Arias, T. A.; McEuen, P. L. *Nature* **2004,** 431, (7006), 284-287.
36. Mounet, N.; Marzari, N. *Phys. Rev. B* **2005,** 71, (20), 205214.

**Figure captions**

**Fig. 1 Drum-type graphene/MoS$_2$ MR.** (a) Schematic of G/MoS$_2$-MR. (b) SEM image of G/MoS$_2$-MR. Raman spectrum for the drum area of G/MoS$_2$-MR for (c) MoS$_2$ and (d) graphene related peaks.

**Fig. 2 Electrical measurement of the atomically thin 2D MR.** (a) Schematic of the measurement setup for the mechanical resonance properties. (b) DC transfer characteristic of the suspended G/MoS$_2$ FET. Frequency response curves measured under various $V_{gs}$'s for (c) G-MR and (d) G/MoS$_2$-MR. Baseline of $I_{MOD\_AM}$ is shifted artificially. Small arrows in the respective curves are the resonance frequencies under each $V_{gs}$.

**Fig. 3 DC gate bias dependence of the resonance frequency.** (a) and (b) DC gate bias dependence of the resonance frequency measured under various temperatures above room temperature (296 K). Solid lines are the fitting curves for eq. 5.

**Fig. 4 Temperature dependence of the resonance frequency shift.** (a) and (b) Temperature dependence of resonance frequency shift $\Delta f_0/f_0$ measured under various $V_{gs}$. Solid lines are the fitting curves to eq. 5. (c) $V_{gs}$ dependences of the temperature coefficients of the resonance frequency shift for G-MR and G/MoS$_2$-MR.

**Table I.** Parameters used for the numerical calculations

| $d_0$ (μm) | $d_t$ (μm) | $E_{gra}$ (TPa) | $E_{MoS2}$ (TPa) | $n_{gra}$ | $n_{MoS2}$ | $\rho_{gra}$ (kg/m$^3$) | $\rho_{MoS2}$ (kg/m$^3$) | $t_{gra}$ (nm) | $t_{MoS2}$ (nm) |
|---|---|---|---|---|---|---|---|---|---|
| 6 | 0.3 | 1.0 | 0.3 | 1 | 3 | 2250 | 5060 | 0.34 | 0.67 |

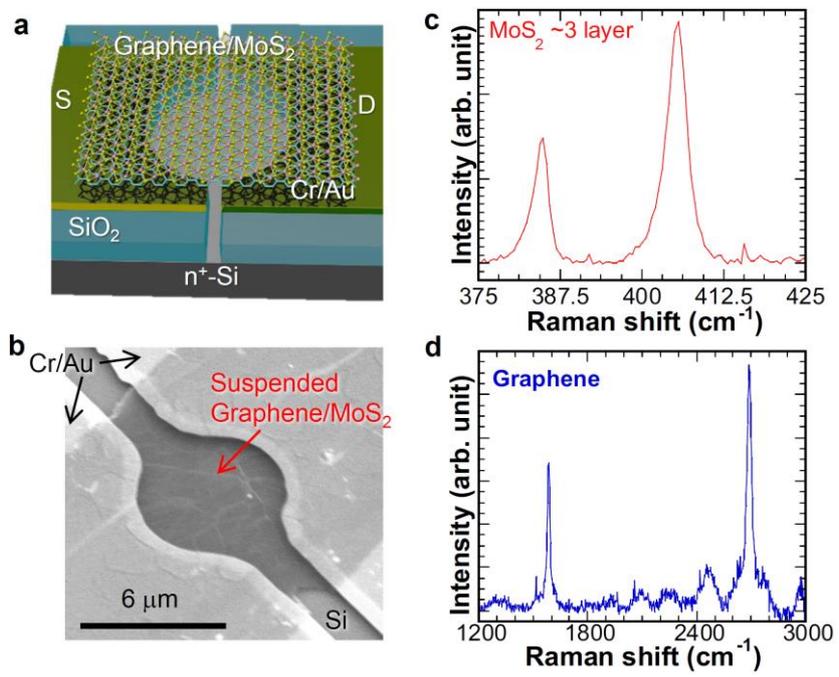

Fig. 1

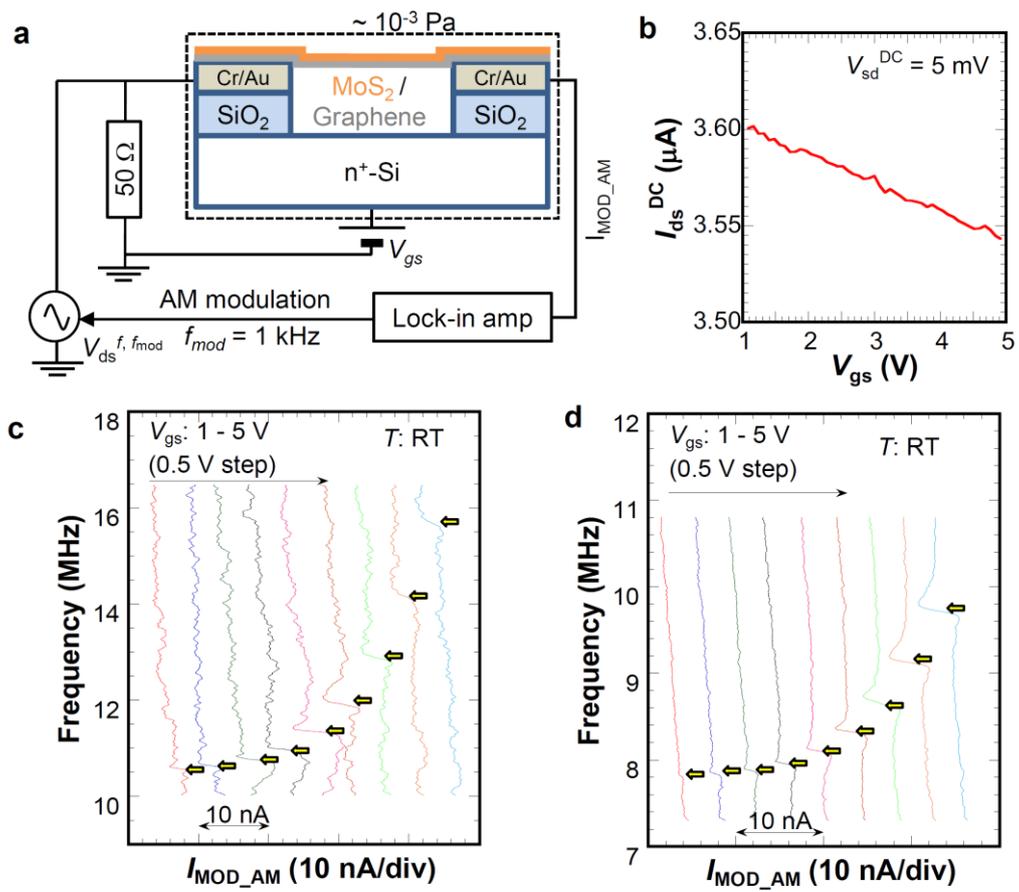

Fig. 2

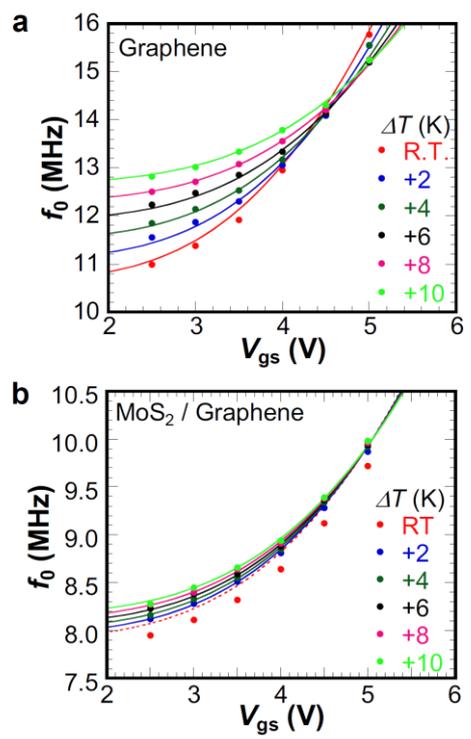

Fig. 3

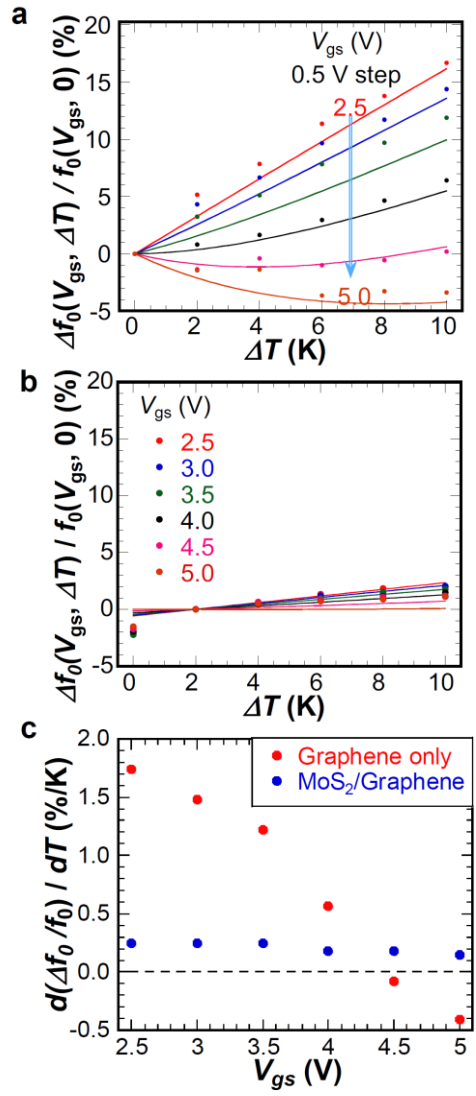

Fig. 3